\documentclass[twocolumn]{proc}
\usepackage{algorithmic}
\usepackage{algorithm}

\def\3{${}'$}

\begin{document}


\author{Tomas Rokicki}
\title{Twenty-Five Moves Suffice for Rubik's Cube}
\date{24 March 2008}
\maketitle
\begin{abstract}
How many moves does it take to solve Rubik's Cube?  Positions are
known that require 20 moves, and it has already been shown that
there are no positions that require 27 or more moves; this is a
surprisingly large gap.  This paper describes a program that
is able to find solutions of length 20 or less at a rate of more
than 16 million positions a second.  We use this program, along
with some new ideas and incremental improvements in other
techniques, to show that there is no position that requires 26
moves.
\end{abstract}

\section{Introduction}

The Rubik's Cube is a simple, inexpensive puzzle with only
a handful of moving parts, yet some of its simplest properties remain
unknown more than thirty years after its introduction.  One of the
most fundamental questions remains unsolved:  how many moves are
required to solve it in the worst case?  We consider a single move
to be a turn of any face, 90 degrees or 180 degrees in any direction
(the `face turn metric').  In this metric, there are more than
36,000 distinct positions known that require at least twenty moves
to solve\cite{rok:pos20}.
No positions are yet known that require twenty-one moves.
The best theoretical approaches and computer searches to date
have only been able to prove there are no positions that require more than
twenty-six moves\cite{ck:2007}.

In this paper, we prove that all positions can
be solved in twenty-five or fewer moves.
We prove this new result by separating the cube space into
two billion sets, each with 20 billion elements.  We then
divide our attention between finding an upper bound on the
distance of positions in specific sets, and combining those
results to calculate an upper bound on the full cube space.

The new contributions of this paper are:
\begin{enumerate}
\item We extend Kociemba's near-optimal solving algorithm to consider
six transformations of a particular position simultaneously, so it
finds near-optimal positions more quickly.
\item We convert his solving algorithm into a set
solver that solves billions of positions at a time.
\item We show how to eliminate a large number of the sets from
consideration, because the positions in them only occur in conjunction
with positions from other sets.
\item We combine the three contributions above with some simple
greedy algorithms to pick sets to solve, and with a fair amount of
computer power, we actually run the sets, combine the results, and
prove that every position in the cube can be solved in 25 moves or
less.
\end{enumerate}

\section{Colors, Moves, and the Size of Cube Space}

The Rubik's cube appears as a stack of 27 smaller cubes (cubies),
with each face of the cubies colored one of six colors.  Of these 27
cubies, seven form a fixed frame around which the other twenty
move.  The seven that form the fixed frame are the center cubies on
each face and the central cubie.

Each move on the cube consists of grabbing the nine cubies that
form a full face of the larger cube, and rotating them as a group
90 or 180 degrees around a central axis shared by the main cube
and the nine cubies.  Each move maintains the set of fully-visible
cubie faces.  The eight corner cubies each always have the same set of
three faces visible, and the twelve edge cubies each always have the
same set of two faces visible.  We will frequently use the term
`corner' to mean `corner cubie', and `edge' to mean `edge cubie'.

In the solved position, each face of the main cube has a single
color.  By convention, we associate these colors with their orientation
on the solved cube:  U(p), F(ront), R(ight), D(own), B(ack), and L(eft).
Each move that uses a a 90 degree clockwise twist is specified by just
specifying the face with no suffix; each move that uses a
90 degree counterclockwise twist is specified with the face followed
by a prime symbol (\3),
and each move that uses a 180 degree twist is specified with the
face followed by the digit 2.
So a clockwise quarter turn of the right face is represented by
R, and the move sequence R2L2U2D2F2B2 generates a pretty pattern
known as {\em Pons Asinorum}.  There are a total of eighteen different
moves; we define this set by $S$.  (By convention, a clockwise turn is
sometimes written with a `+' or `1' suffix and a counterclockwise turn
is sometimes written with a `3' suffix.)

The corner cubies may be permuted arbitrarily, as may the edge
cubies, except that the parity of the permutations of the edge and
corner permutations must match.  This contributes a factor of $12!8!/2$
toward the total number of reachable positions.

Every corner cubie has exactly one face
with either the U or D color.  We define the default orientation
for the corner cubies to be that where the U or D face is on
the whole-cube u or d face; the corner cubies may also be twisted
120 degrees clockwise or counterclockwise with respect to this default
orientation (looking toward the center of the cube).  Note that
these orientations for each cubie are preserved by the moves
U, D, R2, L2, F2, B2, but not by the moves R, L, F, or B.
This corner cubie orientation is fully arbitrary, except
that the sum of all the twists for all the corner cubies must be a
multiple of 360 degrees.  These corner orientations contribute an
additional $3^8/3$ factor toward the total number of reachable positions.

We define the default edge orientation to be that orientation
in the solved state of the cube that is preserved by the moves
U, D, R, L, F2, B2 (but changed by F and B).  Each edge is either
flipped from this orientation or not; the count of flipped edges
must be even.  These edge orientations contribute an additional
$2^{12}/2$ factor toward the total number of reachable positions.

The total number of reachable positions, and thus, the size, of the
cube group, is the product of these factors, which is about 4.33E19.
We call the set of reachable positions $G$.
For each of these positions, an infinite number of move
sequences obtain that position.
We define the {\em distance} of a position $p$ ($d(p)$)
to be the shortest length of any
move sequence that obtains that position.  We define the distance
of a set of positions to be the maximum of the distances of all the
positions in that set.

As a convention, we will denote the successive application of two
move sequences by concatenation.  We will also denote the
application of a sequence to a position, or set of positions,
by concatenation of the position and the sequence.

\section{Symmetry}

The Rubik's cube is highly symmetrical.  There is no distinction
among the faces except for the color; if we were to toss the cube
in the air and catch it, the cube itself remains the same; only
the color corresponding to the u face, the r face, and so on
changes.  Indeed, by tossing the cube, catching it, and noting
the colors on the various faces in the new orientation, we can
enumerate a total of 24 different ways we can orient the cube,
each with a distinct mapping of colors to U, F, R, D, B, and L
faces.
Specifically, there are six different colors the up face can
have, and for each of those six colors, there are four
colors possible for the front face.  These two face colors
fully define the orientation of the normal physical cube.

If we peer in a mirror while performing this experiment, we
notice that our alter-ego holds a cube with mirror-reversed
orientations; these mirror-reversed orientations present an
additional 24 possible mappings from colors to oriented
faces.  We further notice that whenever we do a clockwise move, our
alter ego does a counterclockwise move.

If we choose a canonical color representation, then each
of these 48 orientations is a permutation of the cube colors.
We call this set of color permutations $M$.
If a particular cube position $p$ is obtained by a move
sequence $s$, we can obtain fully corresponding positions
by applying one of the 48 color permutations (say, $m$),
performing the sequence $s$, and then applying the inverse
permutation of $m$.  The resulting position shares many
properties with the original one (especially, for us,
distance).  If we repeat this operation for all 48
permutations in $M$, we will obtain 48 positions.  These positions
are not always unique, but for the vast majority of cube positions
they will be.  Using this form of symmetry, we can reduce many
explorations of the cube space by a factor of 48.

Each cube position has a single specific inverse position.
If a position is reached by a move sequence $s$, then the
inverse position is reached by inverse move sequence $s'$.
To invert a move sequence, you reverse it and invert each
move; the face remains the same, but clockwise becomes counterclockwise
and vice versa.
The set of symmetrical positions of the inverse
of position $p$ is the same as the inverses of the symmetrical
positions of $p$.  Some properties of a position are shared by
its inverse position (specifically, distance).

We can partition the cube space into symmetry-plus-inverse
reduced sets by combining each position with
its symmetrical positions and their inverses; there are only
4.51E17 such sets.

\section{Calculating the Diameter}

We are primarily interested in finding the maximum of the distance
for all positions; this is known as the diameter of the group.
One technique for this is to enumerate all positions, and
optimally solve each.  Practical optimal solvers for Rubik's cube have
been available for some time\cite{korf:1997};
they typically average about 15 minutes
for each optimal solution (with great variance both among programs
and among positions).  If we were to use such a program to solve the
reduced set of 4.51E17 positions, with today's hardware, we would
require more than three million computers for more than three million
years.

We know, however, that some positions require at least twenty moves.
The first such position found is called {\em superflip}; it has every
cubie in the correct place, all corners correctly oriented, and all
edges flipped\cite{reid:1995b}.
Herbert Kociemba devised an algorithm to quickly
find reasonably short but not necessarily optimal solutions to
arbitrary positions.  That
program (slightly improved as we shall describe)
can find move sequences of length twenty or less at a rate of
about 240 positions per second (subject to the condition that there
is such a sequence; no exceptions have been found yet).  Even with
this kind of speed, proving all 4.51E17 positions would require
more than seven thousand computers for more than seven thousand years.

All hope is not lost.  Technology marches onward; when we get to
the point we can solve nine hundred million positions a second, we will need
only four computers for four years to finish the proof.
In the meantime, we can come up with better techniques to refine the
upper bound, and improve our techniques.

\section{Kociemba's Algorithm}

Several techniques have been used to find an upper bound on the diameter
of the cube group.  Thistlewaite gave a four-stage algorithm that
requires a maximum of 52 moves.  Herbert Kociemba improved this to an
algorithm that requires a maximum of 29 moves (as shown by
Michael Reid\cite{reid:1995a}).  Our work is based on
Kociemba's algorithm, so we will describe it a bit further here.
Kociemba himself has a much more detailed explanation on his web
site\cite{koc:exp}.  In 2006, Silviu Radu reduced the upper bound
to 27\cite{radu:2006}, and in 2007 Kunkle and Cooperman reduced
it to 26\cite{ck:2007}.

Kociemba's algorithm identifies a subset of 20 billion positions,
called $H$.  Reid showed that every position in this subset is solvable
in at most 18 moves, and further that every cube position is at most 12
moves from this subset.  Phase one finds a move sequence that takes an
arbitrary cube position to some position in the subset $H$, and phase
two finds a move sequence that takes this new position to the fully
solved state.

To describe this subset, we will introduce some new terminology.
A cubie {\em belongs} in a particular place, if it is in that place in
the solved cube.  Thus, all cubies that have some face colored
u belong in one of the top nine cubies.
The middle layer consists of the nine cubies between the top
and bottom layers; only four of these cubies (edges all) move.

The subset $H$ is composed of all patterns that have the following
characteristics:

\begin{enumerate}
\item All corners and edges are properly oriented.
\item The edge cubies that belong in the middle layer are located
in the middle layer.
\end{enumerate}

The number of positions for which these conditions hold are
the permissible permutations of the corners, the top and bottom edges,
and the middle edges, with the condition that the parity between
the edge permutation and the corner permutation must match.  This is
thus $8!8!4!/2$ or 19.5 billion positions.

These characteristics are preserved by the moves U, U2, U\3, D, D2, D\3,
R2, L2, F2, B2, which we call the set $A$.
Further, these moves suffice to transform every position in $H$ to
the solved state.  (This is a nontrivial result, but it can easily be
shown by brute force enumeration.) For almost all positions
in $H$, the shortest move sequence consisting only of moves from $A$
is the same length as the shortest move sequence consisting only
of moves from $S$, as shown in Table~1.
Further, the worst case is 18 in both cases.

\begin{table}\begin{center}
\begin{tabular}{rrr}
 d&  moves in $S$&  moves in $A$\\ \hline
 0&             1&             1\\
 1&            10&            10\\
 2&            67&            67\\
 3&           456&           456\\
 4&         3,079&         3,079\\
 5&        20,076&        19,948\\
 6&       125,218&       123,074\\
 7&       756,092&       736,850\\
 8&     4,331,124&     4,185,118\\
 9&    23,639,531&    22,630,733\\
10&   122,749,840&   116,767,872\\
11&   582,017,108&   552,538,680\\
12& 2,278,215,506& 2,176,344,160\\
13& 5,790,841,966& 5,627,785,188\\
14& 7,240,785,011& 7,172,925,794\\
15& 3,319,565,322& 3,608,731,814\\
16&   145,107,245&   224,058,996\\
17&       271,112&     1,575,608\\
18&            36&         1,352\\ \hline
  &19,508,428,800&19,508,428,800\\
\end{tabular}
\caption{The number of positions in $H$ at a given distance
using moves from $S$ and moves from $A$; the numbers are
strikingly similar.}
\end{center}\end{table}

Because the defining characteristics of this set treat the U and D
faces differently than the L, R, F, and B faces, all 48 symmetries
of the cube cannot be used; however, 16 can be used.
Thus, after reducing by symmetry, and using
two bits per entry, it is possible to store a distance table for
the entire set $H$ in only about 300MB of memory.  To determine a
sequence that takes a position in $H$ to the solved position,
simply look up the distance for the current position in the large
table.  If it is not zero, try each of the 10 moves in $A$ to find a
position that is closer, and make that move.  Repeat until the cube
is solved.

The remaining problem is how we can transform an arbitrary cube
position into a position in $H$ in 12 or fewer moves.  To illustrate
how this can be done, we describe a way to relabel the cube
so that all positions in $H$ have the same appearance, and all
positions not in $H$ have a different appearance.

Consider an arbitrary position $p$.
To be in $H$, the permutations of the corners are irrelevant;
only the orientation matters.  To represent this, we remove all
colored stickers from the corners, replacing the stickers
colored U or D with U and leaving the other faces, say,
the underlying black plastic.  (To make it easy to follow, we also
replace the D sticker in the center of the bottom face with U.)
All corner cubies are now interchangeable, but we have sufficient
information to note the orientation of the corners.

The permutation of the middle edges does not matter either, but they
must lie in the middle layer and be oriented correctly.  We thus
remove the colored stickers from four edge cubies that belong in
the middle layer, replacing the F and B colors with F and leaving
the L and R colors as black.  (We also replace the B center sticker
with F for convenience.)

The permutations of the top and bottom edges also does not matter;
for these we do the same color change we did for the corners (U and
D get turned into U, and the other four colors get removed).

With this transformation, all positions in $H$ get turned into the
same solved cube:  eight corners, each with a U sticker on either
the up or down face; four middle edges, each with a F sticker on
either the front or back face; eight top/bottom edges, each with
a U sticker on the top or bottom face.  Every position not in $H$
has a different appearance.

This relabeled puzzle has a much smaller state space than the
full cube space.  Specifically, the space consists of
$3^8/3$ corner orientations multiplied by
$2^{12}/2$ edge orientations multiplied by $12\choose 4$ ways to
distribute four middle edges among twelve edge positions, for
a total of 2.22E9 positions.  We call this set of positions $R$.
With 16 ways to reduce this by
symmetry and using only two bits per state, a full distance table
is easy to fit in memory, and the full state space can be explored
easily.  We shall call this relabeling process $r$; it takes a
position in $G$ and transforms it into a position in $R$.

Kociemba's algorithm, then, is to take the original position,
call it $p$, compute $r(p)$, the relabeling; solve the relabeled
puzzle with some sequence $a\in S^*$, apply those moves to an
original cube yielding $pa$ which lies in $H$, and then finish the
solution with another sequence $b\in A^*$ such that $pab$ is
the solved cube.  The final solution sequence is $ab$.

Kociemba's algorithm splits the problem into two roughly equal
subproblems, each of which is easy to exhaustively explore,
using a lookup table that fits in memory,
yielding a fairly good solution to the much larger problem.
This algorithm can find a
solution of distance 29 or less almost instantaneously (in well
under a millisecond).  This defines an upper bound on the
worst-case position distance.

Kociemba extended this algorithm for another purpose: to quickly
find near-optimal solutions for a given position.
He proposed finding many phase 1 solutions, starting with the shortest and
increasing in length, and for each finding the shortest phase
2 solution.  By considering dozens, thousands, or even millions
of such sequences, he has found that in practice nearly optimal
solutions are found very quickly.  Given an input which is the
initial cube position denoted by $p$, his algorithm is given as
Algorithm~1.
The algorithm can either run to completion, or it can be
terminated by the user or when a solution of a desired length
is attained.
\begin{algorithm}
\caption{Kociemba's Algorithm.}
\begin{algorithmic}[1]
   \STATE $d \leftarrow 0$
   \STATE $b \leftarrow \infty$
   \WHILE{$d<b$}
      \FOR{$s \in S^d, r(ps)=e$}
         \IF{$d+d_2(ps)<b$}
            \STATE Solve phase 2; report new better solution
            \STATE $b = d+d_2(ps)$
         \ENDIF
      \ENDFOR
      \STATE $d\leftarrow d + 1$
   \ENDWHILE
\end{algorithmic}
\end{algorithm}

In Kociemba's algorithm, $d_2$ is a table lookup that
takes a position in $H$ and returns the distance to the
identity element ($e$) using moves in $S$.  (Kociemba
actually uses a smaller, faster table that gives a
bound on this value; see \cite{koc:exp} for details.)  The for loop
is implemented by a depth-first recursive routine that
maintains $ps$ incrementally and has a number of further
refinements, such as not permitting $s$ to end in a move
in $A$.  The phase two solution process is omitted both
because it is straightforward and because
it takes much less time than enumerating phase one
solutions.

This algorithm is extremely effective.  Some reasons are:

\begin{enumerate}
\item  Phase one solutions are found very fast, and
mostly access the portions of the phase 1 lookup table near
the solved position; this locality enhances the utility of
caches significantly.
\item  When searching for a phase 2 solution, almost
always the very first lookup shows that the distance to the
solved position would make the total solution longer than
the best found so far; thus, almost all phase 1 solutions are
rejected with a single lookup in the phase 2 table.
\item  Kociemba has found that in practice the algorithm
runs considerably faster if he does not consider phase 1
solutions that contain a strict prefix that is also a phase 1
solution.  This is motivated by the fact that we had already
explored that prefix earlier (since we consider phase 1
solutions by increasing length).
\item  The last move at the end of phase 1 is always
a quarter turn of F, B, R, or L; the inverse move is also a
solution of phase 1, so candidate solutions are always found
in pairs at the leaves of the phase 1 search tree.
\item  There are a number of optimizations that can be
performed for the phase 1 search when the distance to $H$ is
small, such as storing specifically which moves decrease
the distance from that point.
\end{enumerate}

Kociemba's algorithm can be run as described above, or it can
be run in triple-axis mode.  Note how the algorithm treats
the u and d faces differently than the other four.  Instead
of just exploring a single given position $p$, in triple-axis
mode we explore three rotated positions, one with the cube
rotated such that the r and l faces correspond to u and d,
one such that the b and f faces correspond to u and d, and the
original unrotated position.  We try each rotation for a given
phase 1 depth before moving on to the next phase 1 depth.
Our tests show that this finds smaller positions much faster
than the standard single-axis mode; when trying to find solutions
of length 20 or less, this works approximately six times faster
on average than single-axis search.

We have taken this idea one step further; we also consider
the inverse position in three orientations for a new six-axis
mode.  We find this gives on average a further factor of two
speed increase when trying to find positions of twenty moves
or less.

\section{Our Set Solver}

Reid showed a bound of 30 by proving it takes no more than 12 moves to
bring an arbitrary cube position to the $H$ set (by solving
the restickered cube), and then showing that every cube position
in $H$ can be solved in 18 moves.  (He then reduced that to 29
with a clever insight we omit for brevity\cite{reid:1995a}.)
Our proof of 25 is similar,
but instead of using just the $H$ set, we use a union of
thousands of sets all related to $H$.

Consider how Kociemba's solver solves an arbitrary position to
find a near-optimal solution.  It first brings the position
into $H$, using some sequence of moves, and then calculates
within $H$ how close to solved it is.  It then finds another
way to bring the position into $H$, and checks how close it is
to solved at that point.  It does this dozens, or hundreds, or
thousands, millions, or even billions of times, each time
checking for a shorter solution.

We turn this technique inside out.  Instead of using a
big table containing, for each position within $H$, how close it
is to solved, instead we use a table that indicates whether we've
gotten to that particular position in $H$ already.  We then
enumerate all sequences that take that initial position into
$H$, for longer and longer positions, until that table is
completely marked; that every position has been seen.  At this
point, the length of the longest sequence we considered is the
maximum distance from solved for an entire large set of positions;
specifically, all those positions that, when phase-one relabeled, give
the same labeling as the initial position we started with.

With this technique, we are essentially solving 20 billion
cube positions, rather than one; if we were to write out each
sequence every time we set a new bit in the large table, we would
eventually write out an optimal sequence for every position in that set.

The main input to our set solver is a sequence $a\in S^*$, which takes
the solved cube into some position; the set that will be
solved is $Ha$.  Another input is the maximum depth $m$ to run
the phase one search; we have found the value $m=16$ is usually
sufficient to prove an upper bound for the distance of the set to be 20.
To find the exact distance, $m$ should be set to $\infty$.
Our algorithm is given as Algorithm~2.
\begin{algorithm}
\caption{Set Solver}
\begin{algorithmic}[1]
   \STATE $f \leftarrow \emptyset$
   \STATE $d \leftarrow 0$
   \LOOP
      \STATE $f = f \cup Af$ \COMMENT{\it ---prepass}
      \IF{$f = Ha$}
         \STATE {\bf return} $d$
      \ENDIF
      \IF{$d \leq m$}
         \FOR[{\it --search}]{$s \in S^d, r(as)=e$}
            \STATE $f \leftarrow f \cup s^{-1}$
         \ENDFOR
      \ENDIF
      \IF{$f = Ha$}
         \STATE {\bf return} $d$
      \ENDIF
      \STATE $d\leftarrow d + 1$
   \ENDLOOP
\end{algorithmic}
\end{algorithm}
In line 4, we use left multiplication of a set by a move; this is
unconventional but still quite fast.
In line 9, $s$ is the solution to the position reached by
$s^{-1}$.  Unlike Kociemba's search, we do permit our phase 1 search to
enter and then leave the $H$ group; we do this in order to
compute the precise set bound (should we want to).  We have
not yet explored the performance impact of this on our
running time.

The set $f$ is represented by a bitmap, one bit per position.
For the prepass (line 4), we need to have both a source
and destination set, so we need to have two of these bitmaps
in memory at once.  Our memory requirements are completely
dominated by these bitmaps, requiring a total of
4.7GB of memory.  For positions in $R$ that have symmetry, we
take advantage of that symmetry to run the positions on
older machines without that much memory.

The indexing of $f$ is done by splitting the cube position
into independent coordinates, representing the permutation of
the corners, the permutation of the up/down edges,
and finally the permutation of the middle edges.

The time spent in the code is split between the prepass
and the search phases.  The prepass is a simple scan over
the entire $f$ set, multiplying by the ten moves
in $A$; this can be done efficiently by handling the
coordinates from most significant to least significant
in a recursive algorithm so that the inner loop only need
deal with the permutation of the middle edges, and the
more expensive corner coordinate computation is performed
early in the recursion and thus substantially fewer times.

The time in the search phase (lines 9--11) is very
small for low $d$, because
there are few sequences $s$ that satisfy the conditions, but
as $d$ grows, so does the time for the search phase,
exponentially.  Typically a search at level $d+1$ will require
ten times as much time as a search at level $d$.  By limiting
$m$ to 16 in the typical case, we limit the total time in the
search phase, and the whole program runs
fast.  For values of $m$ of 17 or higher, the search phase
will dominate the total runtime.

Our current implementation of this set solver has a major
restriction; at the moment
it only solves those sets for which the middle four edges are
placed in the middle.  This restriction simplifies some of the
move calculations, but it is not essential to our basic ideas.
In the future we plan to lift this restriction.

\section{Improving the Bound}

Some sets we solve have relatively few positions in the
furthest distance.  Since for lower values of $m$ our
set solver only gives us an upper bound on the set
distance, in many cases the true distance of all these
positions is less than the calculated upper bound.
By solving these explicitly using a single-cube solver,
and proving they do not require
as many moves as our set solver found,
we can frequently reduce our bound on
the distance for the set by 1.  To facilitate this,
if the count of unsolved positions in one of the sets
falls below 65,536 at the top of the loop, we print each
of these positions to a log file.

To solve these positions, we first use our six-axis
implementation of Kociemba's
solution algorithm.  Since the solution distance we seek is almost always
19 or 20, this algorithm finds solutions very quickly, usually
in a fraction of a second.  For those positions that
resist Kociemba's solver, we solve them using our optimal
solver.

\section{The Set Graph}

The set $R$ of relabeled positions of $G$ has about
two billion elements.  Consider a position $a\in R$; we can
define the set of $a$ to be all elements $g\in G$ such that
$r(g)=a$.  Let us pick a single one of the elements $i$ in the
set of $a$; the entire set can be represented by $Hi$.
Each set has precisely the same number of elements, about
20 billion; every pair of sets is either identical or disjoint;
and the union of all of the sets is $G$, the full cube space.
(This can be shown with elementary group theory
because $H$ is a subgroup of $G$ and each set $Hi$ is a coset.)

These sets are all related by the moves in $S$.  Consider a
cube position $a$ and its set $Ha$.  The set $Hab$ for
$b\in S$ is adjacent to the set $Ha$.
We can consider $R$ as a graph, where the vertices are the
sets represented by the positions of $R$, and the edges are
moves in $S$.
Clearly for any given position $|d(ab)-d(a)|\le 1$,
and therefore the same is true for sets as a whole:
$|d(Hab) - d(Ha)|\le 1$.  If we have shown that $d(Ha)\leq c$ for
some value of $c$, we have also shown that $d(Has)\leq c+|s|$ where
$s\in S^*$ and $|s|$ is the length of $s$.
This allows us to find an upper bound for one
set, and use it to infer constraints on upper bounds of
neighboring sets in the graph of $R$.

The relabeled puzzle shows 16-way symmetry, so there are really 
only about 139 million relabeled positions when reduced by this
symmetry.  This reduced graph easily fits into memory, and
operations on this graph can be performed reasonably quickly.
For each vertex, we maintain a value which is the least upper
bound we have proved to date.  These values are initialized to
30, since we know every position and thus every set has a distance
of no more than that.  As we
solve new sets, we update the value for the vertex associated
with that set, and update adjacent vertices recursively with the
new upper bound implied by this value.

\section{Eliminating Unneeded Sets}

Just as the triple-axis search improves the performance of
Kociemba's algorithm, we can use remappings to decrease
the amount of work needed to prove a new bound on the diameter.

The R-relabeled cube's space can be defined by three
coordinates:  the placement of the four middle edges, the
orientation of the eight corners, and the orientation of
the twelve edges.  Note that the middle edges are those
between the up and down faces.  If you reorient the cube
so the R and L faces are up and down, then the four
edges that were between R and L are now the middle edges.
Similarly, if you reorient the cube so the F and B faces
are up and down, the the four edges that were between
F and B are now middle edges.  Thus, the same original
cube position has three relabels depending on orientation,
and the middle edge positioning for these three orientations
cover all twelve edges of the cube.

Each corner cubie in the graph is adjacent to three edges.
Therefore, there are eight corners adjacent to an odd
number of edges.  Consider any subset $a$ of edges, and count
the number of corners adjacent to an odd number of edges
in that subset; call that $f(a)$.  For any two subsets
$a$ and $b$, $f(a+b)\le f(a)+f(b)$.  If we partition the
twelve edges into three subsets $a$, $b$, and $c$, and we
know that $f(a)+f(b)+f(c)=8$, then at least one of
$f(a)$, $f(b)$, or $f(c)$ must be three or more.

Therefore, every cube position has some orientation where
the middle edges in that orientation have an odd corner
count ($f$) of three or more.
If we find an orientation-independent property that
applies to all of these subsets, we can extend that property to the
full set of cube positions, since there is some orientation
of every cube position that lies within one of these subsets.

This idea is easily generalized.  The relabeled cube position
information can be described by three independent coordinates:
the middle edge positions, the edge orientations, and the
corner orientations.  For each of coordinates, the full set of
possible combination three-tuples can be explicitly
enumerated.  Given these three-tuples, a covering set based
on any particular weighting can be chosen.  Using the
middle edge position or the corner orientation, it is not
difficult to eliminate a third of the sets; with the
edge orientation, it is not difficult to eliminate half of
the sets.  For the results in this paper, we have eliminated
sets to solve based on the edge positioning, because the
restrictions of our set solver make this the most advantageous
coordinate to use.  We have eliminated 94 possible middle edge
positions out of the 495 possible, including ones that are the
furthest from the middle-edges-in-middle restriction of our
set solver.  This eliminated 25.7 million of the 138.6 million
sets, especially those furthest from the ones that our set
solver could solve.

\section{Choosing Sets to Solve}

This work grew out of a search for distance 21 positions\cite{rok:pos20}
that involved solving a number of these sets exactly.  We thus
started this work with a few thousand cosets already solved; we
used those as our base set.  At every point during this exploration
we maintained the symmetry-reduced graph $R$ on disk annotated with
the best upper bound we had proven for each corresponding set.
To select a new set to solve, we used a simple greedy strategy.
We chose a vertex that, when we pushed its bound down to 20,
and propogated its implications over the graph $R$, it would reduce
the maximum number of vertices from above 25 to 25 or less;
we call this value the `impact' of the vertex.
We annotated each node with this value as well.
The candidate set was small, since restrictions
in our set solver mean that
it can only solve approximately 288,000 of the 139 million sets.

Once we had selected a vertex, we added it to the list of sets
to solve, updated the relevant vertices on the in-memory copy of
the graph (not the persistent one on disk), and repeated this
process to select another vertex.  Since the potential impact
of solving a particular vertex only decreased as other vertices
were solved, we skipped vertices that had a prior impact score
less than the current best; this allowed subsequent vertices to
be selected relatively quickly.

We typically generated lists of a few hundred sets to solve in this
manner.  Since some of the sets actually came in with a bound of
19 or even 18, and this changed the graph in ways differently
than our above algorithm assumed, we generated a brand new list of
vertices to solve every few days based on the updated $R$ graph.

\section{Correctness}

For computer searches such as this one, it is imperative that
the programs be correct.  In order to ensure this, we have
taken a number of measures:

\begin{enumerate}
\item We developed an independent, simple set solver
that uses the simplest possible code with a simple hashtable structure
to store the found positions.  We compared the results of our
fast set solver against this simpler, slower program, to the search
depth that the simpler program was able to handle.  We also
compared the full set of positions found between the two programs
to the depth possible.
\item All set solutions were solved only on server-class machines
containing error-checked-and-corrected memory.
\item All separately solved cube positions were taken as sequences,
and a separate program was run to aggregate these positions into a
simple database of solved positions.  Where upper bounds were
reduced by calculating solutions for small sets of positions, these
sequences were queried and verified that they did, indeed,
yield the appropriate position.
\item Every optimization, including symmetry reductions of various
kinds, search restrictions, parallel multicore operation, and so on,
were switchable on and off, and the results of the program validated.
\item As a sanity check, from the 21 found distance-18 sets,
a random 3,000,000 cube positions were selected and for each of these,
using the six-axis Kociemba solver, a solution of length 18 or better was
found.
\end{enumerate}

\section{Results}

Solving sets with symmetry can be done with machines that do
not have much memory.  Initially our largest machine had
3.5GB of RAM, not enough to solve sets lacking symmetry.
So we started with the symmetrical sets.  The symmetrical
sets are also faster to solve, but the results from the
symmetrical sets were insufficient
to prove a bound of 25.  We then purchased a machine with 8GB of
memory and set it to solving sets lacking symmetry.
We selected sets to solve based on how many neighbors would be
reduced to below 25 by solving that set, assuming an upper
bound of 20.  After solving approximately two thousand
additional sets, we were able to bring the upper bound for
each vertex in the set graph below 26, proving that every
position of Rubik's cube can be solved in 25 or fewer moves.

We have to date found 21 sets that have a distance of 18 or
less, 948 that have a distance of 19 or less, and 7,990 that
have a distance of 20 or less.  Because we initially
focused on symmetrical sets (due to memory constraints),
many of these sets are neighbors, and some are redundant (for
instance, a set with a lower bound of 19, that is adjacent
to a set with a lower bound of 18, is redundant because the
latter implies the former.)  The 21 sets we have found that
have a distance of 18 are listed in Table~2.  These,
and all other sets, are given by a representative element;
$\epsilon$ means the representative element is the empty
sequence so that set is just $H$.

\begin{table}\begin{center}
\begin{tabular}{rl}
d&sequence\\ \hline
18&$\epsilon$\\
18&R\3B2UB2U\3R\\
18&R\3DL2D\3R\\
18&R\3U\3R2F2R\3F2UR\3\\
18&B\3U\3R2UR2B\3\\
18&F\3R2UR2U\3F\3R2U\3\\
18&F\3DFR2U\3FU2F\3\\
18&F\3D\3BR2BDF\\
18&B\3L2R2F\3D2U2F\3U\\
18&B\3U\3F2UB\3U\3\\
18&F\3D\3BR2BDF\3\\
18&B\3D\3L2DL2B\3R2\\
18&R\3F2U2RF2R2U2R\3\\
18&R\3DL2D\3RU\3\\
18&B\3UF2U2F2UB\3\\
18&L\3D\3U\3L\3DUL\3U\3\\
18&F\3UFUF\3U2F\3\\
18&F\3DUF\3D\3U\3F\3U\3\\
18&B\3DUB\3D\3U\3B\3\\
18&F\3D\3F\3R2B\3D\3B\3\\
18&B\3D2R2U2L2F\\
\end{tabular}
\caption{All known sets (by representative) that have a distance
of exactly 18.}
\end{center}\end{table}

Any single set that we've shown to be 18, with the exception of
the $\epsilon$ set immediately
proves a bound of 29 on $G$ because every node in the graph of $R$
is within 11 edges of some orientation of each of these sets.

Proving these sets to be 18 took four or five hours
each; in some cases, we ran our solver with $m=17$, and in other
cases we simply solved the remaining positions at the top level.

No single set we have solved by itself shows a bound of 28 for the
diameter of $G$, but
many pairs of the ones listed above do.  One example such pair is
B\3UF2U2F2UB\3 and B\3U\3R2UR2B\3.  That is, just solving these
two sets proves a bound of 28, because every set in the graph of
R is within 10 moves of some orientation of one of these two sets.

\begin{table}\begin{center}
\begin{tabular}{rl}
d&sequence\\ \hline
20&R\3B2U\3L\3F\3D2R2D\3LB\3U\3\\
20&R\3D2B\3L\3ULF\3R\3F2D\3B\\
19&F\3L\3DF\3U\3R\3B2R\3\\
19&L\3BF\3DU\3B\3RF2L\3F\3U\3\\
\end{tabular}
\caption{Four sets, with the given upper bounds on distances,
that can be added to the set of 18's to prove a distance of 27
for the set $G$.}
\end{center}\end{table}

To prove a bound of 27 for $G$, we include all 21 of the sets
we list above, plus an additional four sets listed in Table~3.
These 25 sets are sufficient to bring every necessary
node of $R$ down to a bound of 27.  This list was determined with
a greedy search.  We started with our set of 21 distance-18 sets,
and propogated their implications over the graph of $R$.  We
then iteratively chose the solved set that had the greatest impact
on $R$ (reduced the maximum number of vertices from 28 or more to
below 28, in this case), until all vertices were below 28.  All
subsequent set counts mentioned were generated by this sort of post-pass
over our collection of solved sets.

To reproduce the result of Cooperman and Kunkle, all of the above
sets plus an additional 127 set bounds that we have found are needed,
for a total of 152 sets.

Our new result requires a total of 3,868 set bounds (about half of
the total number we have solved).  This includes 621 of the sets
for which we've shown a bound of 19, and 3,226 of the sets for
which we've shown a bound of 20.

It is difficult to say what the total execution time has been,
because most of the sets were run slowly, using an older, slower
version of the set solver, on older, slower hardware.  But given
that the current version of the program, on a modern Q6600 CPU
running at 1.6GHz, using a value of $m=16$, and proving a set to
a depth of 20, can be done in 18 minutes on average, and taking
into account the individual positions we solved to lower some of
the set bounds, we believe the total execution time would be on
the order of 1,500 CPU hours to reproduce our proof of 25.

We continue to execute sets, and we are making progress toward
proving a bound of 24.  In order to complete this proof without
requiring an inordinate number of sets, we need
to lift the restriction on our set solver that it only solve sets
that have the middle edges in place.  Once this is done, we
believe that with only a few more CPU months, we can show a new
bound of 24 on the diameter of the cube group.

\section{Acknowledgements}

This work was greatly helped by discussions with Silviu Radu; it
was he who directed us to the subgroup (called $H$ here)
used by Kociemba.  We are also grateful to Herbert Kociemba for
both his original 1992 algorithm (and its implementation in
Cube Explorer) and for some email discussions.

\end{document}